\documentclass[12pt]{article}
\usepackage{graphicx}
\def \app{D_{\pi \pi}}
\def \b{{\cal B}}

\def \bea{\begin{eqnarray}}
\def \beq{\begin{equation}}
\def \bg{\bar \Gamma}
\def \bl{\bar \lambda}

\def \cn{Collaboration}

\def \eea{\end{eqnarray}}
\def \eeq{\end{equation}}
\def \ite{{\it et al.}}

\def \ok{\overline{K}^0}

\def \half{\frac{1}{2}}
\def \3half{\frac{3}{2}}

\begin{document}

\begin{flushright}
EFI 03-14 \\
hep-ph/0304178 \\
April 2003 \\
\end{flushright}

\renewcommand{\thesection}{\Roman{section}}
\renewcommand{\thetable}{\Roman{table}}
\centerline{\bf I-SPIN, U-SPIN, AND PENGUIN DOMINANCE IN $B \to K K \bar K$
\footnote{To be submitted to Physics Letters B.}}
\medskip
\centerline{Michael Gronau\footnote{Permanent Address: Physics Department,
Technion -- Israel Institute of Technology, 32000 Haifa, Israel.}
and Jonathan L. Rosner}
\centerline{\it Enrico Fermi Institute and Department of Physics}
\centerline{\it University of Chicago, Chicago, Illinois 60637}
\bigskip

\begin{quote}

Isospin and U-spin symmetries are useful approximations for studying 
penguin dominance in $B$ meson decays to three kaons, $B\to KK\bar K$. 
We point out certain subtleties in treating these decays in these 
approximations.  Resulting uncertainties are discussed in determining the CP
content of the final state in $B \to K^+K^-K_S$, and in relating the CP
asymmetry in this process to the value of $\sin 2\beta$.

\end{quote}

\leftline{\qquad PACS codes:  12.15.Hh, 12.15.Ji, 13.25.Hw, 14.40.Nd}

\bigskip
Recently the Belle collaboration reported branching ratio measurements 
for $B$ meson decays to three kaon states, $K^+K^- K^+,~K^+K^-K_S,~K_SK_SK^+$
and $K_SK_SK_S$ \cite{Belle1}, and for $B^+$ decays to  $K^+K^-\pi^+$
and $K^+\pi^+\pi^-$ \cite{Belle2}. The BaBar collaboration 
measured branching ratios for $B^+$ decays to all three body final states 
involving charged kaons or pions, $K^+K^-K^+,~K^+K^-\pi^+,~K^+\pi^+\pi^-$
and $\pi^+\pi^+\pi^-$ \cite{BaBar}. The averaged values of the measured 
branching ratios are, in units of $10^{-6}$:
\bea\label{1}
\b(B^+ \to K^+K^+K^-) & = & 30.8 \pm 2.1~~,~~~~\b(B^0 \to K^+K^0K^-) =  
29.3 \pm 5.3~~,\\
\label{2}
\b(B^+ \to K^+K_SK_S) & = & 13.4 \pm 2.4~~,~~~~\b(B^0 \to K_SK_SK_S) = 
4.3 \pm 1.7~~,\\
\label{3}
\b(B^+ \to K^+\pi^+\pi^-) & = & 58.4 \pm 4.4~~,~~~~\b(B^+ \to 
\pi^+\pi^+\pi^-) = 10.9 \pm 3.7~~,\\
\label{4}
\b(B^+ \to K^+K^-\pi^+) & < &  6.3~(90\%~{\rm c.l.})~~.
\eea 
These are averaged over the process shown and its CP-conjugate.

Assuming penguin dominance in $B \to KK\bar K$, an isospin analysis was 
attempted by 
the Belle collaboration \cite{Belle1} in order to isolate the CP-even and 
CP-odd components of the $K^+K^-K_S$ final state. This information is 
useful for studying the 
time-dependent CP asymmetry in this channel \cite{Belle3}. Similar isospin
arguments were presented subsequently in \cite{GLNQ}, where a U-spin 
study relating three body $B^+$ decays involving charged kaons and pions
was employed in order to estimate deviations from penguin dominance. 

In the present Letter we will iterate the isospin analysis for 
$B \to KK\bar K$, pointing out a subtlety 
which was overlooked by the above two studies, thereby oversimplifying the 
analysis. It will be shown that these earlier studies made an 
implicit assumption which goes beyond isospin symmetry. We will argue
that, nevertheless, in the pure penguin limit an equality holds between the 
amplitudes of $B^0 \to K^+K^-K^0$ and $B^+\to K^0\ok K^+$, which is the 
basis for the CP argument. In order to
study deviations from penguin dominance in these decays we will  
employ U-spin considerations which were oversimplified in \cite{GLNQ}. 
We will argue that these deviations, which are partly due to electroweak
penguin contributions, may be larger than estimated, and introduce a sizable 
uncertainty in the value of $\sin 2\beta$ determined from the CP asymmetry 
in $B^0\to K^+K^-K_S$.

The effective Hamiltonian describing charmless decays $B \to KK\bar K$
consists of operators transforming as a sum of $\Delta I = 0$ and 
$\Delta I = 1$. The initial state is pure $|I=\half\rangle$, while the final 
state $|f\rangle \equiv
|KK\bar K\rangle$ is a superposition of three isospin states, $|I(KK)=0,
I_f = \half\rangle,~|I(KK)=1,~I_f=\half\rangle$ and 
$|I(KK)=1,~I_f=\3half\rangle$. Consequently, these decay 
processes are described by five independent isospin amplitudes 
\cite{comment1} corresponding to given $KK\bar K$ momenta.
Suppressing the momentum dependence, we denote these amplitudes by 
$A_{\Delta I}^{I(KK),I_f} \equiv \langle I(KK),~I_f|\,\Delta I\,
|\half\rangle$, and list them as $A_0^{0,\half}$,~$A_0^{1,\half}$,
$A_1^{0,\half}$,~$A_1^{1,\half}$ and $A_1^{1,\3half}$.

$B$ mesons decay into two kaons and an antikaon in four distinct flavor modes, 
$B^+\to K^+K^+K^-$, $B^0\to K^0K^0\ok$, $B^+\to K^+ K^0\ok$ and 
$B^0\to K^+K^0 K^-$. The first two processes involve identical kaons 
($K^+$ and $K^0$, respectively), while in the other two processes the kaons 
($K^+$ and $K^0$ in both decays) have different charges. The amplitudes for $B
\to KK\bar K$ depend on the kaons' momenta. In 
$B^0 \to K^+K^0K^-$ and in $B^+ \to K^+K^0 \ok$ one 
expects different values for amplitudes when exchanging the $K^+$ and $K^0$ 
momenta. Using isospin, one can decompose decay amplitudes for 
$B\to KK\bar K$ into isospin amplitudes $A_{\Delta I}^{I(KK),I_f}$ 
describing two kaons and an antikaon with given momenta. Thus, one obtains 
expressions for six decay amplitudes in terms 
of five isospin amplitudes,   
\bea\label{iso1}
A(K^+K^+K^-)_{p_1p_2p_3} & = & 2A_0^{1,\half} - 2A_1^{1,\half} + 
A_1^{1,\3half}~~,\\ 
\label{iso2}
A(K^0K^0\ok)_{p_1p_2p_3} & = & - 2A_0^{1,\half} - 2A_1^{1,\half} + 
A_1^{1,\3half}~~,\\ 
\label{iso3}
A(K^+K^0\ok)_{p_1p_2p_3} & = & A_0^{0,\half} - A_0^{1,\half} - 
A_1^{0,\half} + A_1^{1,\half} + A_1^{1,\3half}~~,\\ 
\label{iso4}
A(K^+K^0\ok)_{p_2p_1p_3} & = & - A_0^{0,\half} - A_0^{1,\half} + 
A_1^{0,\half} + A_1^{1,\half} + A_1^{1,\3half}~~,\\
\label{iso5}
A(K^+K^0K^-)_{p_1p_2p_3} & = & A_0^{0,\half} + A_0^{1,\half} + 
A_1^{0,\half} + A_1^{1,\half} + A_1^{1,\3half}~~,\\
\label{iso6}
A(K^+K^0K^-)_{p_2p_1p_3} & = & - A_0^{0,\half} + A_0^{1,\half} - 
A_1^{0,\half} + A_1^{1,\half} + A_1^{1,\3half}~~.
\eea
On the left-hand side amplitudes are specified by the three outgoing particles 
and by their respective momenta. On the right-hand side we have absorbed 
Clebsch-Gordan coefficients in the definition of isospin amplitudes and  
have suppressed the momentum dependence of these amplitudes. 

Let us comment briefly on Eqs.~(\ref{iso1})--(\ref{iso6}).
The equal magnitudes and the relative signs of contributions of isospin 
amplitudes in pairs of processes can be easily understood from simple 
considerations. In Eqs.~(\ref{iso1}) and (\ref{iso2}) the final states 
involve $I(KK)=1$ and are related to each other by an isospin reflection 
$u \leftrightarrow d,~\bar d \leftrightarrow -\bar u$. Consequently, the 
magnitudes of $\Delta I = 0$ and $\Delta I=1$ contributions in these two 
processes are equal and occur with opposite and equal signs, respectively. 
In Eq.~(\ref{iso3}) and (\ref{iso4}) one interchanges the $K^+$ and $K^0$ 
momenta in $B^+\to K^+K^0\ok$. It then follows from Bose statistics that 
contributions to the two amplitudes from $I(KK) = 0$ terms, which are 
antisymmetric in the isospins of the two kaons, and $I(KK) = 1$ terms, 
which are symmetric in isospin, are equal and have opposite and equal 
signs, respectively. 
This corresponds to situations in which the orbital angular momentum of 
the $KK$ system in its center of mass frame (which equals the $\bar K$ 
angular momentum relative to this center of mass) is odd and even, 
respectively. 
The same argument applies to Eqs.~(\ref{iso5}) and 
(\ref{iso6}), giving the amplitude for $B^0 \to K^+K^0K^-$ in two points
of phase space where the two kaon momenta are interchanged.

An interesting consequence of the isospin decomposition is a sum rule
between $B^+$ and $B^0$ decay amplitudes. The six amplitudes 
(\ref{iso1})--(\ref{iso6}) obey one linear relation between the sum of three 
amplitudes for a charged $B$ meson and the sum of three amplitudes for a 
neutral $B$,
\bea\label{SR}
A(K^+K^+K^-)_{p_1p_2p_3} + A(K^+K^0\ok)_{p_1p_2p_3}
+ A(K^+K^0\ok)_{p_2p_1p_3} & = & \nonumber\\
A(K^0K^0\ok)_{p_1p_2p_3}
+ A(K^+K^0K^-)_{p_1p_2p_3} + A(K^+K^0K^-)_{p_2p_1p_3} & = & 
3A_1^{1,\3half}~~.
\eea
The two sums, in each of which one sums over amplitudes at two points of
phase space where $K^+$ and $K^0$ momenta are interchanged, are given by the 
$I_f = \3half$ amplitude.
This relation is similar to an isospin relation among the four amplitudes
for $B^+$ and $B^0$ decays to $K\pi$ \cite{IsoKpi}. 

So far our arguments were based purely on isospin symmetry. Let us now study
the consequences of penguin dominance in $B\to KK\bar K$ decays,
assuming that the dominant term in the $\Delta C= 0,~\Delta S =1$ effective 
Hamiltonian \cite{BBL} contributing to these decays is a $\bar b \to \bar s$ 
QCD penguin operator.  
The assumption of penguin dominance implies that one keeps only 
$\Delta I = 0$ terms in Eqs.~(\ref{iso1})--(\ref{iso6}), neglecting 
$A_1^{0,\half}$ $A_1^{1,\half}$ and $A_1^{1,\3half}$. Note that this
excludes electroweak penguin operators which contain a term transforming 
as $\Delta I = 1$. We will return to this point when 
discussing deviations from penguin dominance. In the latter approximation 
one has \bea\label{A1}
A(K^+K^+K^-)_{p_1p_2p_3} & = & -A(K^0K^0\ok)_{p_1p_2p_3} = 2A_0^{1,\half}~~,
\\
\label{A0+1}
A(K^+K^0K^-)_{p_1p_2p_3} & = & -A(K^0K^+\ok)_{p_1p_2p_3} =
A_0^{0,\half} + A_0^{1,\half}~~,
\eea
where $A_0^{0,\half}$ and $A_0^{1,\half}$ are antisymmetric and symmetric 
under interchange of the two kaon momenta.  
The equality of the two pairs of $B^+$ and $B^0$ decay amplitudes follows 
simply from an isospin reflection $u \leftrightarrow d$ in
initial and final states. In this limit the two amplitudes involving $K^+K^-$
in the final state are, however, different.

When squaring the amplitudes and integrating over phase 
space one includes a factor $\half$ for identical particles in the first 
pair of processes. The interference between the two isospin amplitudes
$A_0^{0,\half}$ and $A_0^{1,\half}$ in the second pair of processes, 
corresponding to even and odd angular momenta of the $KK$ system, 
vanishes. Thus, one finds
\bea\label{rate1}
\Gamma(B^+\to K^+K^+K^-) & = & \Gamma(B^0 \to K^0K^0\ok)
= 2\Gamma_0^{1,\half}~~,\\
\label{rate2}
\Gamma(B^0 \to K^+K^0K^-) & = & \Gamma(B^+ \to K^+K^0\ok) = 
\Gamma _0^{0,\half} + \Gamma_0^{1,\half}~~,
\eea
where $\Gamma_0^{0,\half}$ and $\Gamma_0^{1,\half}$ are rates 
corresponding to the two isospin amplitudes. We conclude that, while the two
rate equalities (\ref{rate1}) and (\ref{rate2}) follow from penguin dominance, 
not all four rates are equal in this approximation. In particular, the two 
rates involving $K^+K^-$ in the final state 
may be different in general, contrary to arguments made in 
\cite{Belle1,GLNQ}. They become equal when $\Gamma _0^{0,\half} = 
\Gamma_0^{1,\half}$ which goes beyond isospin symmetry. Experimentally, 
one has $\Gamma(B^+\to K^+K^+K^-)/\Gamma(B^0 \to K^+K^0K^-) = 0.98 
\pm 0.19$ which implies $\Gamma_0^{0,\half}/\Gamma_0^{1,\half} = 
1.05^{+0.49}_{-0.33}$. 

Eq.~(\ref{A0+1}) provides the basis for attempting a separation between 
the CP-even and CP-odd components in the final state of the measured 
process $B^0\to K^+K^-K_S$ \cite{Belle1}. These components correspond 
to even and odd angular momentum $K^+K^-$ states. Equal amplitudes in 
$B^0\to K^+K^0K^-$ and $B^+\to K^+K^0\ok$ imply that the final states 
in the two processes have the same angular momentum decomposition in 
terms of $K^+K^-$ in one process and $K^0\ok$ in the other. The 
probability for a $K^0\ok$ being in an even angular momentum state, where 
it decays as $K_SK_S + K_LK_L$, is given by 
$2\Gamma(B^+ \to K^+K_SK_S)/\Gamma(B^+\to K^+K^0\ok)$. Using 
Eq.~(\ref{rate2}), this probability is given by a ratio of two 
measured rates $2\Gamma(B^+ \to K^+K_SK_S)/\Gamma(B^0\to K^+K^0K^-) =
1.04 \pm 0.20$, excluding the $\phi K_S$ contribution in the denominator
\cite{Belle1}. This is also the probability for a CP-even state in 
$B^0\to K^+K^-K_S$ excluding $\phi K_S$.

The above conclusion, indicating that the final state in $B^0\to K^+K^-K_S$
(excluding $\phi K_S$) is dominantly CP-even, is based on assuming penguin 
dominance in $B\to KK\bar K$. The rest of the discussion will address 
this issue. We will study tests for penguin dominance in $B\to KK\bar K$, 
and will evaluate deviations from this approximation in terms of 
measurable rates. 

Penguin dominance in 
$B \to K\pi$, suggested in \cite{GHLR}, was first tested in \cite{DGR}
by using flavor SU(3) and comparing decay rates for $B\to K\pi$ and 
$B\to \pi\pi$. The measured rates were also used to estimate the deviation 
from pure penguin dominance, given by a parameter $|T'/P'| \sim 0.2$.
A similar analysis will be presented here in order to relate 
$B\to KK\bar K$ to $B\to\pi\pi\pi$ and $B\to K\pi\pi$ to $B\to KK\pi$. 
This will test penguin dominance in $B\to KK\bar K$.    
Our arguments differ in detail from those presented in 
\cite{Belle1,GLNQ}. In \cite{Belle1} factorization was assumed for three 
body decays for which no good theoretical justification exists, while in 
\cite{GLNQ} a subtlety in using U-spin was overlooked. We will also 
explain the special role of electroweak penguin contributions which 
were ignored in the latter study.

A useful subgroup of SU(3) permitting relations between $B^+$
meson decays to final states involving charged pions and kaons is U-spin
\cite{Uspin}, under which the pairs $(d,s),~(\bar s,-\bar d),~(\pi^-,K^-)$ 
and $(K^+,-\pi^+)$ transform like doublets. The $\Delta C=0,~\Delta S = 1$ 
effective Hamiltonian transforms like a $\bar s$ component 
($\Delta U_3 = \half$) of a U-spin doublet, while the $\Delta C=0,
\Delta S = 0$ Hamiltonian transforms like a $\bar d$ component 
($\Delta U_3 = -\half$) of another U-spin doublet.
Let us consider the decays of $B^+$ into the final states $K^+K^+K^-,
K^+\pi^+K^-,~K^+\pi^+\pi^-$ and $\pi^+\pi^+\pi^-$. The initial state in 
these decays is pure $U=0$. The final states from the 
$\Delta U =\half$ transitions are two U-spin doublets, in which the  
two positively charged particles are in $U(++) = 0$ and $U(++) = 1$ 
states. Therefore, $\Delta S=1$ and $\Delta S=0$ amplitudes may be written  
separately in terms of two U-spin amplitudes corresponding to these two 
states \cite{comment2}. We will denote $\Delta S=1$
amplitudes corresponding to $U(++) = 0$ and $U(++) = 1$ by $A^0_s$ 
and $A^1_s$,
respectively, and analogous $\Delta S=0$ amplitudes by $A^0_d$ and $A^1_d$.

Absorbing Clebsch-Gordan coefficients in the definition of these amplitudes
and specifying the three outgoing particle momenta, one finds
\bea\label{U1}
A(K^+K^+K^-)_{p_1p_2p_3} & = & 2A^1_s~~,\\
\label{U2}
A(K^+\pi^+\pi^-)_{p_1p_2p_3} & = & -A^0_s + A^1_s~~,\\
\label{U3}
A(K^+\pi^+\pi^-)_{p_2p_1p_3} & = & A^0_s + A^1_s~~,\\
\label{U4}  
A(\pi^+\pi^+\pi^-)_{p_1p_2p_3} & = & 2A^1_d~~,\\
\label{U5}
A(K^+\pi^+K^-)_{p_1p_2p_3} & = & A^0_d + A^1_d~~,\\
\label{U6}
A(K^+\pi^+K^-)_{p_2p_1p_3} & = & -A^0_d + A^1_d~~.
\eea
The similar forms of corresponding $\Delta S=1$ and $\Delta S=0$ 
amplitudes may be easily understood in terms of a simple U-spin 
reflection $d \leftrightarrow s$. The relative signs of terms in 
amplitudes in which two momenta are interchanged follow from Bose 
statistics. 

Two amplitude relations follow from U-spin,
\bea
A(K^+\pi^+\pi^-)_{p_1p_2p_3} + A(K^+\pi^+\pi^-)_{p_2p_1p_3} & = &
A(K^+K^+K^-)_{p_1p_2p_3}~~,\\
A(K^+\pi^+K^-)_{p_1p_2p_3} + A(K^+\pi^+K^-)_{p_2p_1p_3} & = &
A(\pi^+\pi^+\pi^-)_{p_1p_2p_3}~~.
\eea
On the left-hand-side one sums over amplitudes at two points in phase space 
where the $K^+$ and $\pi^+$ momenta are interchanged.

Squaring Eqs.~(\ref{U1})--(\ref{U6}) and integrating over phase space, 
one obtains
\bea\label{Gamma_s}
\Gamma(B^+\to K^+K^+K^-) & = & 2\Gamma^1_s~~,~~~~~~~
\Gamma(B^+\to K^+\pi^+\pi^-) = \Gamma^0_s + \Gamma^1_s~~,\\
\label{Gamma_d}
\Gamma(B^+\to \pi^+\pi^+\pi^-) & = & 2\Gamma^1_d~~,~~~~~~~
\Gamma(B^+\to K^+\pi^+K^-) = \Gamma^0_d + \Gamma^1_d~~,
\eea
where $\Gamma^0_{s,d}$ and $\Gamma^1_{s,d}$ are rates corresponding to the 
two U-spin amplitudes. Contrary to arguments presented in \cite{GLNQ},
this does not imply an equality between $\Gamma(B^+\to K^+K^+K^-)$ and 
$\Gamma(B^+\to K^+\pi^+\pi^-)$, or between 
$\Gamma(B^+\to \pi^+\pi^+\pi^-)$ and $\Gamma(B^+\to K^+\pi^+K^-)$.
We conclude that U-spin predictions cannot be tested in simple
rate equalities. Instead, as shown in \cite{Uspin}, U-spin 
predicts equal CP rate differences between all pairs of U-spin related 
decays. For instance, the CP rate differences in $B^+\to K^+K^+K^-$ and 
$B^+\to \pi^+\pi^+\pi^-$ are equal in the U-spin symmetry limit. 
Experimental tests of such predictions are quite challenging. 

In order to relate $\Delta S=1$ and $\Delta S =0$ processes to each other,
we decompose the corresponding effective Hamiltonians into terms 
multiplying given CKM factors \cite{Uspin}, 
\bea\label{Hs}
{\cal H}_{\rm eff}^{\bar b\to\bar s} & = & V^*_{ub}V_{us}O^s_u + 
V^*_{cb}V_{cs}O^s_c~~,\\
\label{Hd}
{\cal H}_{\rm eff}^{\bar b\to\bar d} & = & V^*_{ub}V_{ud}O^d_u + 
V^*_{cb}V_{cd}O^d_c~~,
\eea
where $O^{d,s}_u$ and $O^{d,s}_c$ are two operators transforming like 
U-spin doublets. This implies 
\bea\label{As}
A^{0,1}_s & = & V^*_{ub}V_{us}A^{0,1}_u + V^*_{cb}V_{cs}A^{0,1}_c~~,\\
\label{Ad}
A^{0,1}_d & = & V^*_{ub}V_{ud}A^{0,1}_u + V^*_{cb}V_{cd}A^{0,1}_c~~,
\eea
where $A^{0,1}_u$ and $A^{0,1}_c$ are matrix elements of the U-spin doublet 
operators $O_u$ and $O_c$, respectively, for U-spin final states
with $U(++) = 0,1$. 

Eqs.~(\ref{As}) and (\ref{Ad}) may be used to test an assumption of
penguin and electroweak penguin dominance in strangeness changing decays, 
namely an assumption that the second term in Eq.~(\ref{As}) dominates the 
amplitude. We note, however, that in contrast to the isospin analysis which
assumed QCD penguin dominance, the second term includes also contributions 
from electroweak penguin operators.

Applying Eqs.~(\ref{As}) and (\ref{Ad}) to $B^+\to K^+K^+K^-$ and 
$B^+\to \pi^+\pi^+\pi^-$, respectively, which involve the amplitudes 
$A^1_u$ and $A^1_c$, and defining 
momentum-dependent amplitudes
$|P|e^{i \delta}\equiv V_{cb}^* 
V_{cs}A_c^1,~|T|e^{i \gamma}\equiv  V_{ub}^* V_{us}A_u^1$, one has
\bea\label{TP}
A(B^+ \to K^+ K^- K^+) & = & |T|e^{i \gamma} + |P|e^{i \delta}~~,\\
A(B^+ \to \pi^+ \pi^- \pi^+) & = & |T|\bl^{-1}e^{i \gamma} - 
|P|\bl e^{i \delta}~~~,
\eea
where
\beq
\frac{V_{us}}{V_{ud}} = - \frac{V_{cd}}{V_{cs}} = \bl \equiv
\frac{\lambda}{1 - (\lambda^2/2)} = 0.226~~,~~~~~~(\lambda = 0.22)~~.
\eeq
In the amplitudes for $B^-$ decays to charge conjugate final states the 
weak phase $\gamma$ appears with opposite signs. Thus, one obtains for 
the two charge averaged partial widths:
\bea
\bg(B^{\pm} \to K^{\pm} K^{\pm} K^{\mp}) & = & \int [|T|^2 + |P|^2 +
 2 |T||P| \cos \gamma \cos\delta]~~,\\
\bg(B^{\pm} \to \pi^{\pm} \pi^{\pm} \pi^{\mp}) & = & \int
[|T|^2\bl^{-2} + |P|^2\bl^2 - 2 |T||P| \cos \gamma \cos\delta]~~,
\eea
where $\int$ stands for an integral over three-body phase space.  We seek an
upper bound on
\beq \label{zdef}
z \equiv \left[ \int |T|^2 / \int |P|^2 \right]^{1/2}~~~,
\eeq
the tree-to-penguin amplitude ratio averaged over phase space.

(1) Let us assume that $B^\pm \to\pi^\pm \pi^\pm \pi^\mp$ is dominated by the
tree amplitude and $B^\pm \to K^\pm K^\pm K^\mp$ by the penguin. 
Using the observed value of the ratio
\beq
{\cal R} \equiv \frac{\bg(B^\pm \to\pi^\pm \pi^\pm \pi^\mp)}
{\bg(B^\pm \to K^\pm K^\pm K^\mp)} = 0.35 \pm 0.12~~~,
\eeq
we then find $z = 0.13 \pm 0.02 < 0.16~(90\%~{\rm c.l.})$.  However, the
possibility of tree-penguin interference 
weakens this bound somewhat as we show now.

(2) The Schwarz inequality implies that in the definition
\beq\label{xi}
\int |T||P| \cos \delta \equiv \xi \left[ \int |T|^2 \right]^{1/2}
\left[ \int |P|^2 \right]^{1/2}~~~,
\eeq
the magnitude of the parameter $\xi$ 
cannot exceed 1,
and equals 1 when the penguin and tree amplitudes are proportional to each 
other and are relatively real over the entire phase space. 
One can then show that $z$ is a
monotonically increasing function of $\xi \cos \gamma$ for any realistic value
of ${\cal R}$.  An upper bound $\xi \cos \gamma \le 0.74$ is based on assuming
$\gamma \ge 42^\circ$ \cite{PDG}.  (The bound on $z$ is not very sensitive to
this assumption.)  Since we do not know $\delta$ \cite{delta}, we shall
regard $\cos \delta$ as unrestricted.  One obtains the maximum value of $z$ for
$\xi \cos \gamma = 0.74$: $z = 0.19 \pm 0.03$ or $z < 0.23~(90\%~{\rm c.l.})$.

Let us make a few remarks about deviations from penguin dominance 
which were neglected in Eq.~(\ref{A0+1}):
\begin{enumerate}
\item The bound on the tree-to-penguin amplitude $z$ is based on applying 
U-spin in order to relate $B^+\to K^+K^+K^-$ and $B^+\to\pi^+\pi^+\pi^-$. 
Large U-spin breaking effects are expected to affect a relation between 
these two processes. This may result in a value of $z$ as large as 0.3.
\item Corrections to penguin and electroweak penguin 
dominance affecting
Eq.~(\ref{A0+1}) may differ, in both their flavor and momentum dependence,
from those estimated above for 
$B^+ \to K^+K^+K^-$. Therefore, the bound on $z$ can only 
be used indirectly to set an upper limit of this order on corrections to 
Eq.~(\ref{A0+1}) from tree amplitudes.
We conclude that the tree amplitude contribution to $A(K^+K^0K^-) + 
A(K^0K^+\ok)$ could be as large as 0.3 of the equal (but opposite in
sign) penguin amplitudes contributing to these two processes.
\item Electroweak penguin contributions
affect the amplitude equality (\ref{A0+1}). Such terms,
which were included in the denominator of Eq.~(\ref{zdef}) but not in 
its numerator,
involve the same weak phase as the dominant penguin amplitude.
A rough estimate of electroweak penguin corrections to 
the amplitude equality, based on Wilson coefficients or on model 
calculations \cite{FL}, is about $10-20\%$. 
\end{enumerate} 
The combined correction to Eq.~(\ref{A0+1}) from $\Delta I =1$ tree and 
electroweak penguin amplitudes is hard to calculate, and depends on the 
interference between tree and electroweak penguin amplitudes which may be 
constructive except in special cases \cite{NR}. Constructive interference 
could imply an overall correction of $40-50\%$ in the most pessimistic case. 

The effects of electroweak penguin and tree amplitudes on the time dependent 
CP asymmetry in $B^0 \to K^+K^-K_S$ are of two kinds.
Both amplitudes affect the CP structure of the final state $f=K^+K^-K_S$, 
thereby multiplying by a dilution factor the coefficient $S_f$ of the 
$\sin\Delta mt$ term, where $S_f = \sin 2\beta$ for a purely CP-even state 
excluding $\phi K_S$. One way of determining the 
dilution factor is through a partial wave analysis in the angular momentum 
of the $K^+K^-$ system. This may separate even and odd angular momenta 
corresponding to even and odd CP states. 

The tree amplitude ($T_f$) in $B^0 \to K^+K^-K^0$, which has a weak phase 
different from that of the penguin and electroweak penguin amplitudes 
(combined to $P_f$), has another effect. [We use the same convention for
$T_f$ and $P_f$ as in (\ref{TP})].
It modifies $S_f$ from a value $\sin 2\beta$ for a pure 
CP-even state (excluding $\phi K_S$)
to $\sin 2\beta_{\rm eff},~\beta_{\rm eff} = \beta + 
\delta\beta$, and introduces in the asymmetry a $\cos\Delta mt$ term 
with coefficient $C_f$. These corrections depend on the tree-to-penguin 
ratio of amplitudes and on the relative strong phase between 
these amplitudes ($\delta_f$), both varying over phase space. 
We define 
\beq \label{z_f}
z_f \equiv \left[ \int |T_f|^2 / \int |P_f|^2 \right]^{1/2}~~~,
\eeq
\beq\label{xi_c}
\int |T_f||P_f| \cos \delta_f \equiv \xi_c \left[ \int |T_f|^2 \right]^{1/2}
\left[ \int |P_f|^2 \right]^{1/2}~~,
\eeq
\beq\label{xi_s}
\int |T_f||P_f| \sin \delta_f \equiv \xi_s \left[ \int |T_f|^2 \right]^{1/2}
\left[ \int |P_f|^2 \right]^{1/2}~~.
\eeq

Keeping only linear terms in $z_f$, one 
obtains for $C_f$ and $\delta\beta$ expressions which generalize those 
obtained for two body decays \cite{MG}
\bea
C_f & \simeq & 2z_f\xi_s\sin\gamma~~,\\
\delta\beta & \simeq & z_f\xi_c\sin\gamma~~.
\eea 
We estimated that $z_f$ may be as large as 0.3. The largest possible effect 
on $S_f$ occurs when $|\xi_c|=1$, corresponding to
penguin and tree amplitudes which are proportional to each 
other and are relaively real over the entire phase space. 
In this case $|\delta\beta|$ 
could be as large as $17^\circ$, and $S_f$ could lie in the range $0.2-1$.
The Schwarz inequality, $\xi_c^2 + \xi_s^2 \le 1$, implies  
$\xi_s=0$ when $|\xi_c|=1$, hence $C_f=0$. The other extreme 
but unlikely case, of a maximal $|C_f|=2z_f\sin\gamma$ and a minimal  
$\delta\beta = 0$, occurs when $\xi_s=\pm 1,~\xi_c=0$. This corresponds to 
penguin and tree amplitudes which are proportional to each other and are out 
of phase by $\pi/2$ over the entire phase space.

\medskip
M. G. wishes to thank the CERN Theory Division and the Enrico Fermi Institute
at the University of Chicago for their kind hospitality. 
This work was supported in part by the United States Department of Energy
through Grant No.\ DE FG02 90ER40560.

\def \ajp#1#2#3{Am.\ J. Phys.\ {\bf#1}, #2 (#3)}
\def \apny#1#2#3{Ann.\ Phys.\ (N.Y.) {\bf#1}, #2 (#3)}
\def \app#1#2#3{Acta Phys.\ Polonica {\bf#1}, #2 (#3)}
\def \arnps#1#2#3{Ann.\ Rev.\ Nucl.\ Part.\ Sci.\ {\bf#1}, #2 (#3)}
\def \art{and references therein}
\def \cmts#1#2#3{Comments on Nucl.\ Part.\ Phys.\ {\bf#1}, #2 (#3)}
\def \cn{Collaboration}
\def \cp89{{\it CP Violation,} edited by C. Jarlskog (World Scientific,
Singapore, 1989)}
\def \econf#1#2#3{Electronic Conference Proceedings {\bf#1}, #2 (#3)}
\def \efi{Enrico Fermi Institute Report No.}
\def \epjc#1#2#3{Eur.\ Phys.\ J.\ C {\bf#1}, #2 (#3)}
\def \ib{{\it ibid.}~}
\def \ibj#1#2#3{~{\bf#1}, #2 (#3)}
\def \ijmpa#1#2#3{Int.\ J.\ Mod.\ Phys.\ A {\bf#1}, #2 (#3)}
\def \ite{{\it et al.}}
\def \jhep#1#2#3{JHEP {\bf#1}, #2 (#3)}
\def \jpb#1#2#3{J.\ Phys.\ B {\bf#1}, #2 (#3)}
\def \mpla#1#2#3{Mod.\ Phys.\ Lett.\ A {\bf#1} (#3) #2}
\def \nat#1#2#3{Nature {\bf#1}, #2 (#3)}
\def \nc#1#2#3{Nuovo Cim.\ {\bf#1}, #2 (#3)}
\def \nima#1#2#3{Nucl.\ Instr.\ Meth.\ A {\bf#1}, #2 (#3)}
\def \npb#1#2#3{Nucl.\ Phys.\ B~{\bf#1}, #2 (#3)}
\def \npps#1#2#3{Nucl.\ Phys.\ Proc.\ Suppl.\ {\bf#1}, #2 (#3)}
\def \PDG{Particle Data Group, K. Hagiwara \ite, \prd{66}{010001}{2002}}
\def \pisma#1#2#3#4{Pis'ma Zh.\ Eksp.\ Teor.\ Fiz.\ {\bf#1}, #2 (#3) [JETP
Lett.\ {\bf#1}, #4 (#3)]}
\def \pl#1#2#3{Phys.\ Lett.\ {\bf#1}, #2 (#3)}
\def \pla#1#2#3{Phys.\ Lett.\ A {\bf#1}, #2 (#3)}
\def \plb#1#2#3{Phys.\ Lett.\ B {\bf#1} (#3) #2}
\def \prl#1#2#3{Phys.\ Rev.\ Lett.\ {\bf#1} (#3) #2}
\def \prd#1#2#3{Phys.\ Rev.\ D\ {\bf#1} (#3) #2}
\def \prp#1#2#3{Phys.\ Rep.\ {\bf#1} (#3) #2}
\def \ptp#1#2#3{Prog.\ Theor.\ Phys.\ {\bf#1}, #2 (#3)}
\def \rmp#1#2#3{Rev.\ Mod.\ Phys.\ {\bf#1} (#3) #2}
\def \rp#1{~~~~~\ldots\ldots{\rm rp~}{#1}~~~~~}
\def \si90{25th International Conference on High Energy Physics, Singapore,
Aug. 2-8, 1990}
\def \slc87{{\it Proceedings of the Salt Lake City Meeting} (Division of
Particles and Fields, American Physical Society, Salt Lake City, Utah, 1987),
ed. by C. DeTar and J. S. Ball (World Scientific, Singapore, 1987)}
\def \slac89{{\it Proceedings of the XIVth International Symposium on
Lepton and Photon Interactions,} Stanford, California, 1989, edited by M.
Riordan (World Scientific, Singapore, 1990)}
\def \smass82{{\it Proceedings of the 1982 DPF Summer Study on Elementary
Particle Physics and Future Facilities}, Snowmass, Colorado, edited by R.
Donaldson, R. Gustafson, and F. Paige (World Scientific, Singapore, 1982)}
\def \smass90{{\it Research Directions for the Decade} (Proceedings of the
1990 Summer Study on High Energy Physics, June 25--July 13, Snowmass,
Colorado),
edited by E. L. Berger (World Scientific, Singapore, 1992)}
\def \tasi{{\it Testing the Standard Model} (Proceedings of the 1990
Theoretical Advanced Study Institute in Elementary Particle Physics, Boulder,
Colorado, 3--27 June, 1990), edited by M. Cveti\v{c} and P. Langacker
(World Scientific, Singapore, 1991)}
\def \TASI{{\it TASI-2000:  Flavor Physics for the Millennium}, edited by 
J. L. Rosner (World Scientific, 2001)}
\def \yaf#1#2#3#4{Yad.\ Fiz.\ {\bf#1}, #2 (#3) [Sov.\ J.\ Nucl.\ Phys.\
{\bf #1}, #4 (#3)]}
\def \zhetf#1#2#3#4#5#6{Zh.\ Eksp.\ Teor.\ Fiz.\ {\bf #1}, #2 (#3) [Sov.\
Phys.\ - JETP {\bf #4}, #5 (#6)]}
\def \zpc#1#2#3{Zeit.\ Phys.\ C {\bf#1}, #2 (#3)}
\def \zpd#1#2#3{Zeit.\ Phys.\ D {\bf#1}, #2 (#3)}


\begin{thebibliography}{99}
\bibitem{comment1} The dependence of isospin amplitudes on $I(KK)$ was 
overlooked in \cite{GLNQ} where only three isospin amplitudes were 
identified.
\bibitem{comment2} The dependence of U-spin amplitudes on the U-spin of 
the two positively charged particles was overlooked in \cite{GLNQ}, where 
only one U-spin amplitude was identified.
\bibitem{delta} Factorization calculations, which are questionable for 
three body decays, would imply $\cos\delta \simeq -1$
and hence a more restrictive bound on $z$.
\end{thebibliography}
\end{document}